\documentstyle[11pt]{article}
\begin{document}

\title{A New Kind of Deformed Hermite Polynomials and Its Applications
  \thanks{This project supported by National Natural Science Foundation
   of China and LWTZ 1298.}}
\author{{Sicong Jing and Weimin Yang}\\
{\small Department of Modern Physics, University of Science and}\\
{\small Technology of China, Hefei, Anhui 230026, P.R.China}}
\maketitle
\bigskip
\bigskip
\bigskip
\begin{abstract}
A new kind of deformed calculus was introduced recently in studying of
parabosonic coordinate representation. Based on this deformed calculus, a new
deformation of Hermite polynomials is proposed, its some properties such
as generating function, orthonormality, differential and integral
representaions, and recursion relations are also discussed in this paper.
As its applications, we calculate explicit forms of parabose squeezed number
states, derive a particularly simple subset of minimum uncertainty states for
parabose amplitude-squared squeezing, and discuss their basic squeezing
behaviours.
\end{abstract}

\section{Introduction}
Parastatistics was introduced by Green as an exotic possibility extending the
Bose and Fermi statistics \cite{s1} and for the long period of time the
interest to it was rather academic. Nowadays it finds some applications in the
physics of the quantum Hall effect \cite{s2} and (probably) it is relevant to
high temperature superconductivity \cite{s3}. The paraquantization, carried
out at the level of the algebra of creation and annihilation operators,
involves trilinear(or double) commutation relations in place of the bilinear
relations that characterize Bose and Fermi statistics. Recently, the
trilinear commutation relations of single paraparticle systems was rewritten
as bilinear commutation relations by virtue of the so-called R-deformed
Heisenberg algebra \cite{s4}. For instance, the trilinear commutation
relations \cite{s5}
\begin{equation}
\left[ a, \{ a^\dagger , a \} \right] = 2a, ~~~\left[ a,
\{ a^\dagger, a^\dagger \} \right] = 4a^\dagger, ~~~\left[ a,
\{ a, a \} \right] = 0,
\end{equation}
where $a^\dagger$ and $a$ are parabose creation and annihilation operators
respectively, can be replaced by \cite{s6}
\begin{equation}
[ a, a^\dagger ] = 1 + ( p-1 ) R, ~~~\{ R, a \} =
\{ R, a^\dagger \} = 0, ~~~R^2 = 1,
\end{equation}
where $R$ is a reflection operator and $p$ is the paraquantization order
$(p=1,2,3,...)$. Obviously, the bilinear commutation relations (2) may be
treated as some kind of deformation of the ordinary Bose commutator with
deformation parameter $p$.
\par
>From the experience of studying q-deformed oscillators \cite{s7},
we know that it will
be very useful if one introduces corresponding deformed calculus to analyse
the parabose systems. This was done recently and based on the new deformed
calculus, the parabosonic coordinate representation was developed \cite{s8}.
Since special functions play important roles in mathematical physics, it is
reasonable to imagine that some deformation of the ordinary special functions
based on the new deformed calculus will also play similar roles in studying
the parabose systems. In this paper, we introduce a new kind of deformation
for the ordinary Hermite polynomials and demonstrate its various useful
properties.
\par
It is well-known that squeezed state is one of the most important
non-classical states for the usual bose system. In fact in phase space the
simultaneous measurements of canonical variables $x$ (position) and $P$
(momentum) are restricted to a limit of accuracy due to the Heisenberg
uncertainty principle, one may attempt to measure one component in a
quadrature to much greater accuracy at the cost of high uncertainty in the
other component. This has been realized in the case of squeezed states for
bosons. Since the squeeze operator is identical for both the ordinary and
parabose systems, it is natural to investigate squeezing behaviours of a 
parabose system. Applying the squeeze operator to parabose number states gives
the so-called parabose squeezed number states. We derive explicit forms for
these states and show they can be expressed by virtue of the deformed Hermite
polynomial, whose argument is the parabose creation operator multiplied by a
constant, acting on the vacuum. We discuss basic squeezing properties of these
states. Especially, we show the parabose squeezed vacuum is a minimum
uncertainty state for normal squeezing (i.e., either in $x$ and $P$ direction
squeezing), as well as for amplitude-squared squeezing (i.e., squeezing in
variables that are quadratic of the creation or annihilation operators). It is
also interesting to ask, besides the para-squeezed vacuum, if there exist
other states for a one mode parabose system, which are not only
amplitude-squared squeezing but also minimum uncertainty states? We present a
particularly simple subset of such states, that is, the deformed Hermite
polynomial states. These states may or may not be squeezed in the normal
sense.
\par
The paper is organized as follows. In Section 2, for the sake of
self-contained of the present paper, we briefly mention the basic idea of the
new kind of deformed calculus. The deformed Hermite polynomials are
introduced in section 3, their orthonormality, generating function,
differential and integral representaions, and recursion relations are also
discussed in this section. In section 4, we show that the deformed Hermite
polynomials can be used to give the explicit forms for parabose squeezed
number states, which are constructed by applying the squeeze operator to the
parabose number states, and analyse the basic squeezing properties of these
states. In section 5 by solving an eigenvalue equation that allows one to find
the minimum uncertainty states, we present the subset of its solutions which
has a particularly simple form, and discuss the properties of these states.

\section{Deformed calculus related to parabosonic coordinate representation}
It is well-known that parabose algebra is characterized by the double
commutation relations (1). If one demands that the usual relations
\begin{equation}
a = \frac{x+iP}{\sqrt{2}} ,~~~a^\dagger = \frac{x-iP}{\sqrt{2}}
\end{equation}
still work for the parabose case, where $x$ and $P$ stand for the coordinate
and momentum operator respectively, it can be proved that the most genaral
expression for the momentum operator $P$ in the coordinate $x$ diagonal
representation is of \cite{s5} \cite{s8}
\begin{equation}
P = -i \frac{d}{dx} -i \frac{p-1}{2x} (1-R),
\end{equation}
where $p$ is the paraquantization order and $R$ the reflection operator which
has property $R f(x) = f(-x)$ in the coordinate representation for any $x$
dependent function $f(x)$. From (4) a new derivative operator $D$ can be
defined which acts on function $f(x)$ as
\begin{eqnarray}
D f(x)&\equiv&\frac{D}{Dx} f(x) = \frac{d}{dx} f(x) +
\frac{p-1}{2x} (1-R) f(x) \nonumber\\
&=&d\,f(x) + \frac{p-1}{2x} \left( f(x) - f(-x) \right) ,
\end{eqnarray}
where $d\,f=\frac{d}{d\,x}\,f$. Definition (5) implies that $D$ acts on an
even function $f_e (-x)= f_e (x)$ as the ordinary derivative $Df_e (x)=
d\, f_e (x)$, and $D$ acting on an odd function $f_o (-x)= -f_o (x)$ leads to
$Df_o (x) = d\, f_o (x) + \frac{p-1}{x} f_o (x)$. For $p=1$ case, $D$ reduces
to the ordinary derivative operator $d$. Since $P= -iD$, Eq.(3) means that the
pair $(x,D)$ in realization of parabose algebra for a single degree of
freedom plays the same role as $(x,d)$ in realization of the ordinary Bose
algebra. Like q-deformed calculus in which the q-analogue of the number
system was defined by $[n]_q = \frac{q^n -1}{q-1}$ \cite{s7}, such that when
$q \rightarrow 1$, $[n]_q \rightarrow n$, in the present case, one can
introduce a new kind of deformed number system which is defined by
\begin{equation}
[n] = n + \frac{p-1}{2} (1 - (-)^n ).
\end{equation}
Obviously, $[2k] = 2k$, $[2k+1] = 2k+p$ for any integer $k$ and when $p
\rightarrow 1$, $[n] \rightarrow n$. So paraquantization order $p$ may be
referred to as a deformation parameter. In terms of the number system $[n]$,
basis vectors of Fock space for single mode of parabose oscillators take the
usual form
\begin{equation}
|n \rangle = \frac{(a^\dagger)^n}{\sqrt{[n]!}} |0 \rangle, ~~~
a^\dagger |n \rangle = \sqrt{[n+1]} |n+1 \rangle, ~~~
a |n \rangle = \sqrt{[n]} |n-1 \rangle,
\end{equation}
where $[n]!=[n][n-1]...[1], [0]! \equiv 1$, and $|0 \rangle$ is the unique
vacuum vector satisfying $a |0 \rangle =0, aa^\dagger |0 \rangle = p
|0 \rangle$. Generalization of the ordinary differential relation
$d x^n = nx^{n-1} $ reads
\begin{equation}
D x^n = [n]\,x^{n-1},
\end{equation}
which reveals the effect of the deformed derivative operator $D$ on the
polynomials of $x$. If we introduce a notation $E(x)$ defined by
\begin{equation}
E(x) = \sum_{n=0}^{\infty} \frac{x^n}{[n]!},
\end{equation}
we also have
\begin{equation}
D E(x) = E(x)
\end{equation}
Therefore the $E(x)$ is a deformation of the ordinary exponential function
$e^x$ in our case and it will reduce to $e^x$ when $p \rightarrow 1$.
It is worthy of metion that in some special situation the
usual Leibnitz rule also works for the deformed operator $D$
\begin{equation}
D(f\,g) = (Df)\,g + f\,(Dg),
\end{equation}
where either $f(x)$ or $g(x)$ is an even function of $x$.

Of course, inversion of the deformed derivative operator $D$ also leads to a
new deformed integration which may be formally written as \cite{s8}
\begin{eqnarray}
\!& &\! \int \,Dx F(x) = \sum_{n=0}^{\infty} (-)^n \left( \int \,dx
        \frac{p-1}{2x} (1-R) \right)^n \int \,dx F(x) \nonumber\\
\!&=&\! \int \,dx F(x) - \int \,dx \frac{p-1}{2x} (1-R) \int \,dx F(x)
        \nonumber\\
\!& &\! +\left( \int \,dx \frac{p-1}{2x} (1-R) \right)^2 \int \,dx F(x)
        - \cdots .
\end{eqnarray}
>From this expression, it is easily seen that if $F(x)$ is an odd function of
$x$, its deformed integration will reduce to the ordinary integration, that
is, $\int \,Dx F(x) = \int \,dx F(x)$ for $F(-x)=-F(x)$. Corresponding to Eq.
(8), one has
\begin{equation}
\int \,Dx \,x^n = \frac{x^{n+1}}{[n+1]} + c,
\end{equation}
where c is an integration constant. Eq.(12) gives a formal definition for the
deformed integration in the sence of indefinite integral. For definite
integral, we have
\begin{eqnarray}
\!& &\! \int_a^b \,Dx F(x) = \int_a^b \,dx \sum_{n=0}^{\infty} (-)^n \left(
        \frac{p-1}{2x} (1-R) \int_a^x \,dx \right)^n F(x) \nonumber\\
\!&=&\! \int_a^b \,dx F(x) - \int_a^b \,dx \frac{p-1}{2x} (1-R) \int_a^x \,dx
        F(x)  \nonumber\\
\!& &\! + \int_a^b \,dx \frac{p-1}{2x} (1-R) \int_a^x \,dx
        \frac{p-1}{2x} (1-R) \int_a^x \,dx F(x)  - \cdots .
\end{eqnarray}
If either $F(x)$ or $G(x)$ is an even function of $x$, one has a formula of
integration by parts from Eq.(11)
\begin{equation}
\int_a^b \,Dx \frac{D\,F}{Dx} \,G = FG |_a^b -
\int_a^b \,Dx F\,\frac{D\,G}{Dx}.
\end{equation}

\section{Deformed Hermite polynomials and their properties}
Let us consider solutions of a second-order differential equation based on the
deformed derivative operator $D$ defined in the previous section
\begin{equation}
D^2 f(x) -2xDf(x) + \mu f(x) =0.
\end{equation}
In terms of the ordinary derivative notation, Eq.(16) can be rewritten as
$$
\frac{d^2}{dx^2} f(x) - \left( 2x - \frac{p-1}{x} \right)
\frac{d}{dx} f(x) - (p-1) \left( 1 + \frac{1}{2x^2} \right) f(x)
$$
\begin{equation}
+ (p-1) \left( 1 + \frac{1}{2x^2} \right) f(-x) + \mu f(x) = 0.
\end{equation}
We find out that when the parameter $\mu$ takes eigenvalues $\mu = 2[n],
n=0,1,2,3,...$, for each given paraquantization order $p$, the deformed
second-order differential equation (16) has solutions (eigenfunctions) which
form a set of orthogonal functions in the whole real $x$ axis.
In fact, it is not difficult to see that the following polynomials
\begin{equation}
H_n^{(p)} (x)= [n]! \sum_{k=0}^{[n/2]^{'}} \frac{(-)^k (2x)^{n-2k}}{k!
[n-2k]!}
\end{equation}
are the desired solutions of the deformed second-order differential equation
(16) for $\mu = 2[n]$ which will reduce to the usual Hermite polynomials when
$p \rightarrow 1$, where $[k]^{'}$ in the above of summation notation $\sum$
stands for the largest integer smaller than or equal to $k$. So the
polynomials (18) may be considered as a deformation of the usual Hermite
polynomials. The first few polynomials of $H_n^{(p)} (x)$ have
the following explicit forms
\begin{eqnarray}
& &H_0^{(p)}(x)=1, ~~~H_1^{(p)}(x) =2x, ~~~H_2^{(p)}(x) =4x^2-[2]!,
\nonumber\\
& &H_3^{(p)}(x)=8x^3-4[3]x, ~~~H_4^{(p)}(x) =16x^4-16[3]x^2+2[3]!,
\nonumber\\
& &H_5^{(p)}(x)=32x^5-32[5]x^3+8[5][3]x, \nonumber\\
& &H_6^{(p)}(x)=64x^6-96[5]x^4+48[5][3]x^2-[5]!.
\end{eqnarray}
In order to convince oneself that the polynomials (18) are indeed solutions
of eq.(16), one may substitute (18) into (16) and check coefficients of all
powers of $x$ being zero. To do this, more convenient forms for $H_n^{(p)}
(x)$ are
\begin{equation}
H_{2l}^{(p)}(x)=(-)^l[2l]! \sum_{k=0}^{l} \frac{(-)^k (2x)^{2k}}{(l-k)![2k]!}
\end{equation}
and
\begin{equation}
H_{2l+1}^{(p)}(x)=(-)^l[2l+1]! \sum_{k=0}^{l} \frac{(-)^k (2x)^{2k+1}}{(l-k)!
[2k+1]!},
\end{equation}
where $l$ are non-negative integers. Thus we have
\begin{equation}
D^2 H_n^{(p)} (x) -2xD H_n^{(p)} (x) +2[n]H_n^{(p)} (x)=0,
\end{equation}
or, by virtue of the Leibnitz rule (11), we also have
\begin{equation}
D\left( e^{-x^2}DH_n^{(p)}(x) \right) + 2[n]e^{-x^2} H_n^{(p)}(x) =0.
\end{equation}
Also from Eq.(18) we know that $H_n^{(p)} (-x) = (-)^n H_n^{(p)} (x)$, which
means that the deformed Hermite polynomial $H_n^{(p)} (x)$ has its parity
$(-)^n$.
\par
As in the ordinary Hermite polynomials case, the deformed ones also have their
generating function. Using the deformed exponential function $E(x)$ defined
by (9), we can write the generating function of $H_n^{(p)}(x)$ as
\begin{equation}
e^{-t^2} E(2tx) = \sum_{n=0}^{\infty} \frac{t^n}{[n]!} H_n^{(p)} (x).
\end{equation}
We also would like to point out a differential representation of the deformed
Hermite polynomials $H_n^{(P)} (x)$ being
\begin{equation}
H_n^{(p)} (x) = (-)^n e^{x^2} D^n e^{-x^2}.
\end{equation}
Before writing out an integral representation of $H_n^{(p)}(x)$, let us
introduce a notation $[x+y]^n$ which is defined as
\begin{equation}
[x+y]^n \equiv \sum_{k=0}^{n} \frac{[n]!}{[k]![n-k]!} x^{n-k} y^k.
\end{equation}
Using this notation, the integral representation of $H_n^{(p)}(x)$ can be
written as 
\begin{equation}
H_n^{(p)} (x) = 2^n N_0^2 \int_{-\infty}^{\infty} \,Dt \,[x+it]^n
\,e^{-t^2},
\end{equation}
where $N_0$ is determined by
\begin{equation}
N_0^{-2} = \int_{-\infty}^{\infty} \,Dt \,e^{-t^2}.
\end{equation}
Using the following integral formula \cite{s8}
\begin{equation}
N_0^2 \int_{-\infty}^{\infty} \,Dt \,t^{2n} \,e^{-t^2} = \frac{[1][3] \cdots
[2n-1]}{2^n}, ~~~(n>0)
\end{equation}
it is easily to show that equation (27) works. In fact,
\begin{equation}
2^n N_0^2 \int_{-\infty}^{\infty} \,Dt \,[x+it]^n \,e^{-t^2}
=2^n N_0^2 \sum_{k=0}^{n} \frac{[n]!i^k x^{n-k}}{[k]![n-k]!}
\int_{-\infty}^{\infty} \,Dt \,t^k\,e^{-t^2}.
\end{equation}
Noticing that for odd $k$, the deformed integration in the right-hand side of
(30) will reduce to ordinary integration which has no any contribution to
the summation, the above summation can be written as
\begin{equation}
2^n N_0^2 \sum_{k=0}^{[n/2]^{'}} \frac{(-)^k x^{n-2k}}{[2k]![n-2k]!} \int_
{-\infty}^{\infty} \,Dt \,t^{2k}\,e^{-t^2} = H_n^{(p)} (x),
\end{equation}
where (29) has been used. 
\par
Now we turn to the question of demonstrating the orthonormality of $H_n^{(p)} (x)$
\begin{equation}
\int_{-\infty}^{\infty} \,Dx \,e^{-x^2}\,H_n^{(p)} (x) \,H_m^{(p)} (x)
= \frac{2^n [n]!}{N_0^2} \delta_{n,m}.
\end{equation}
Firstly we show the orthogonality of $H_n^{(p)} (x)$
\begin{equation}
\int_{-\infty}^{\infty} \,Dx \,e^{-x^2}\,H_n^{(p)}(x)\,H_m^{(p)}(x) =0,
 ~~~(n \neq m).
\end{equation}
For $n+m$ odd case, Eq.(33) works obviously. In fact, from the parity of
$H_n^{(p)}(x)$ we have $H_n^{(p)} (-x) H_m^{(p)} (-x) = -H_n^{(p)} (x)
H_m^{(p)} (x)$, which means that the
deformed integration in (33) will reduce to an ordinary integration with an
odd integrand $e^{-x^2}H_n^{(p)} (x) H_m^{(p)} (x)$ over a whole real $x$
axis, therefore the integration should be zero. For $n+m$ even case, because
of (23), $H_n^{(p)} (x)$ and $H_m^{(p)} (x)$ satisfy the following equations
\begin{equation}
D \left( e^{-x^2} D H_n^{(p)} (x) \right) + 2[n] e^{-x^2} H_n^{(p)} (x) =0,
\end{equation}
\begin{equation}
D \left( e^{-x^2} D H_m^{(p)} (x) \right) + 2[m] e^{-x^2} H_m^{(p)} (x) =0
\end{equation}
respectively. Multiplying Eq.(34) and Eq.(35) by $H_m^{(p)} (x)$ and
$H_n^{(p)} (x)$ respectively, and substracting the
resulting equations, then integrating it over the whole real $x$ axis, we have
$$
\int_{-\infty}^{\infty} \,Dx \left( H_m^{(p)} \frac{D}{Dx} \left( e^{-x^2}
\frac{D}{Dx} H_n^{(p)} \right) - H_n^{(p)} \frac{D}{Dx} \left( e^{-x^2}
\frac{D}{Dx} H_m^{(p)} \right) \right)
$$
\begin{equation}
+ 2([n]-[m]) \int_{-\infty}^{\infty} \,Dx e^{-x^2} H_n^{(p)} H_m^{(p)} = 0.
\end{equation}
Since the first integration in Eq.(36) satisfies condition of the deformed
integration by parts either for $n$ and $m$ being all even or all odd, we can
write it as
\begin{eqnarray}
& &\int_{-\infty}^{\infty} \,Dx \frac{D}{Dx} \left( e^{-x^2} (H_m^{(p)}
\frac{D}{Dx} H_n^{(p)} - H_n^{(p)} \frac{D}{Dx} H_m^{(p)} ) \right)
\nonumber\\
&=&e^{-x^2} \left( H_m^{(p)} \frac{D}{Dx} H_n^{(p)} - H_n^{(p)} \frac{D}{Dx}
H_m^{(p)} \right) |_{-\infty}^{\infty} = 0,
\end{eqnarray}
here we have used a fact that $H_n^{(p)} (x)$, $H_m^{(p)} (x)$ and their
deformed derivatives are all polynomials of $x$, so the right-hand side of
(37) equals to zero. Noticing that $n \neq m$, we obtain Eq.(33).
\par
Then we calculute integration
\begin{equation}
I_n = \int_{-\infty}^{\infty} \,Dx e^{-x^2} H_n^{(p)} (x) H_n^{(p)} (x).
\end{equation}
Substituting Eq.(25) into this integration, we have
\begin{equation}
I_n = \int_{-\infty}^{\infty} \,Dx H_n^{(p)} (x) (-)^n \frac{D^n}{Dx^n}
e^{-x^2}.
\end{equation}
Noticing that the integration (38) also satisfies the condition of the
deformed integration by parts no matter what non-negative integer $n$ is,
we get
$$
I_n = (-)^n H_n^{(p)} (x) D^{n-1} e^{-x^2} |_{-\infty}^{\infty} - (-)^n
\int_{-\infty}^{\infty} \,Dx \frac{D\,H_n^{(p)}(x)}{Dx} \frac{D^{n-1}}
{Dx^{n-1}} e^{-x^2}
$$
$$
=-e^{-x^2} H_n^{(p)} (x) H_{n-1}^{(p)} (x)|_{-\infty}^{\infty} - (-)^n
\int_{-\infty}^{\infty} \,Dx \frac{D\,H_n^{(p)}(x)}{Dx} \frac{D^{n-1}}
{Dx^{n-1}} e^{-x^2}.
$$
Continuing the procedure of integration by parts, at last we arrive at
\begin{equation}
I_n = \int_{-\infty}^{\infty} \,Dx \frac{D^n\,H_n^{(p)}(x)}{Dx^n}e^{-x^2}
=\frac{2^n[n]!}{N_0^2}.
\end{equation}
Thus we demonstrated the equation (32).
\par
To conclude this section, let us mention that there are some definite
relations between neighbouring deformed Hermite polynomials and their
derivatives which are called recursion relations of
$H_n^{(p)} (x)$. The main recursion relations are the following two:
\begin{equation}
DH_n^{(p)} (x)-2[n]H_{n-1}^{(p)} (x)=0,
\end{equation}
\begin{equation}
H_{n+1}^{(p)} (x) -2xH_n^{(p)} (x) +2[n]H_{n-1}^{(p)} (x) =0.
\end{equation}
It is straightforward to prove these relations by virtue of the definition
(18) and the deformed differential relation (8).

\section{Parabose squeezed number states}
As an application of the deformed Hermite polynomials $H_n^{(p)} (x)$, let us
consider the resulting states from parabose squeezed operator $S(r)=
e^{\frac{r}{2}a^2-\frac{r}{2}(a^{\dagger})^2}$ acting on the parabose number
states $|n \rangle$
\begin{equation}
|r,n \rangle = S(r)|n\rangle,    ~~~(n=0,1,2,3,\cdots)
\end{equation}
here for the sake of simplicity, we take the squeezing parameter $r$ as a
real number. We call $|r,n \rangle$ the parabose squeezed number states.
Obviously, $|r,n \rangle$ form a complete and orthonormal state-vector set
for the single mode of parabose system:
\begin{equation}
\langle r,n| r,m \rangle =\langle n|m \rangle= \delta_{n,m}, ~~~
\sum_{n=0}^{\infty} |r,n \rangle \langle r,n|=1.
\end{equation}
Using the following transformations
\begin{eqnarray}
S(r)\,a\,S(r)^{-1}&=& cosh r \,a + sinh r \,a^\dagger,      \nonumber\\
S(r)\,a^{\dagger} \,S(r)^{-1}&=& cosh r \,a^\dagger + sinh r \,a
\end{eqnarray}
and the disentangling formula
\begin{equation}
S(r)=exp\left(-\frac{tanh r}{2} \,(a^\dagger)^2 \right) exp\left(-
\frac{\ln cosh r}{2} \,(a^\dagger a+a a^\dagger) \right)
exp\left(\frac{tanh r}{2} \,a^2 \right),
\end{equation}
we can write $|r,n \rangle$ as
\begin{equation}
|r,n \rangle = \frac{(sech r)^{p/2}}{\sqrt{[n]!}} \left( cosh r \,a^\dagger +
sinh r \,a \right)^n e^{-tanh r \,(a^\dagger)^2 /2} |0 \rangle.
\end{equation}
In terms of the deformed Hermite polynomials $H_n^{(p)} (x)$, the explicit
form of $|r,n \rangle$ is
\begin{equation}
|r,n \rangle = \frac{(sech r)^{p/2}}{\sqrt{[n]!}} \left( -\frac{1}{2} tanh r
\right)^{n/2} H_n^{(p)} \left( \frac{a^\dagger}{i \sqrt{sinh 2r}} \right)
e^{-tanh r \,(a^\dagger)^2 /2} |0\rangle.
\end{equation}
We prove (48) by induction. Firstly, from (47), and using
\begin{equation}
e^{tanh r \,(a^\dagger)^2 /2}a^n e^{-tanh r \,(a^\dagger)^2 /2}
= (a - tanh r \,a^\dagger)^n,
\end{equation}
for $n=1$ case, we have
$$
|r,1 \rangle= \frac{(sech r)^{p/2}}{\sqrt{[1]!} cosh r} \,a^\dagger
e^{-tanh r \,(a^\dagger)^2 /2} |0 \rangle
$$
\begin{equation}
=\frac{(sech r)^{p/2}}{\sqrt{[1]!}}
\left(-\frac{1}{2} tanh r\right)^{1/2} H_1^{(p)} (\chi)
e^{-tanh r \,(a^\dagger)^2 /2} |0\rangle,
\end{equation}
where $\chi$ stands for $\frac{a^\dagger}{i \sqrt{sinh 2r}}$. Then supposing
\begin{equation}
|r,n-1 \rangle=\frac{(sech r)^{p/2}}{\sqrt{[n-1]!}} \left( -\frac{1}{2}
tanh r \right)^{\frac{n-1}{2}} H_{n-1}^{(p)} (\chi)
e^{-tanh r \,(a^\dagger)^2 /2} |0\rangle,
\end{equation}
we have
\begin{eqnarray}
&&|r,n \rangle=\frac{1}{\sqrt{[n]}}\left( cosh r \,a^\dagger + sinh r \,a
\right)|r,n-1 \rangle  \nonumber\\
&&=\frac{(sech r)^{p/2}}{\sqrt{[n]!}} \left( -\frac{1}{2} tanh r
\right)^{\frac{n-1}{2}} cosh r \,a^\dagger H_{n-1}^{(p)}
(\chi) \,e^{-tanh r \,(a^\dagger)^2 /2} |0 \rangle \nonumber\\
&& + \frac{(sech r)^{p/2}}{\sqrt{[n]!}} \left( -\frac{1}{2} tanh r
\right)^{\frac{n-1}{2}} sinh r \,e^{-tanh r \,(a^\dagger)^2 /2} (a-tanh r
\,a^\dagger)  \nonumber\\
&& H_{n-1}^{(p)} (\chi) |0 \rangle.
\end{eqnarray}
Using the following relation
\begin{equation}
[a,(a^\dagger)^n]=(a^\dagger)^{n-1} \left(n+\frac{p-1}{2}(1-(-)^n)R \right)
\end{equation}
it is easily to find
\begin{equation}
a\,H_{n-1}^{(p)} (\chi) |0\rangle
=\frac{2[n-1]}{i \sqrt{sinh 2r}}\,H_{n-2}^{(p)}(\chi)|0\rangle.
\end{equation}
Substituting (54) into (52), we obtain
\begin{eqnarray}
|r,n\rangle &=& \frac{(sech r)^{p/2}}{\sqrt{[n]!}} \left(-\frac{1}{2}tanh
r\right)^{n/2}\left(2\chi H_{n-1}^{(p)}(\chi)-2[n-1]H_{n-2}^{(p)}(\chi)
\right)    \nonumber\\
& & e^{-tanh r \,(a^\dagger)^2 /2} |0\rangle.
\end{eqnarray}
Thus (48) is proved by virtue of the recursion relation (42).
\par
>From (3) we have Hermitian operators $x$ and $P$ defined by
\begin{equation}
x=\frac{a+a^\dagger}{\sqrt{2}}, ~~~P=\frac{a-a^\dagger}{i\sqrt{2}},
\end{equation}
which satisfy the commutator $[x,P]=i[a,a^\dagger]=i(1+(p-1)R)$. The variances
of the operators $x$ and $P$ in the parabose squeezed number states are of the
form
\begin {eqnarray}
\langle r,n|(\triangle x)^2|r,n \rangle
&\equiv& \langle r,n|x^2|r,n \rangle-\langle r,n|x|r,n
\rangle^2=e^{-2r}(n+\frac{p}{2}),    \nonumber\\
\langle r,n|(\triangle P)^2|r,n \rangle
&\equiv& \langle r,n|P^2|r,n \rangle-\langle r,n|P|r,n
\rangle^2=e^{2r}(n+\frac{p}{2}),
\end{eqnarray}
which lead to
\begin{equation}
\langle r,n|(\triangle x)^2|r,n \rangle \langle r,n|(\triangle P)^2|r,n
\rangle  = \left(n+ \frac{p}{2}\right)^2.
\end{equation}
On the other hand, according to the uncertainty relation
\begin{equation}
\langle r,n|(\triangle x)^2|r,n \rangle \langle r,n| (\triangle P)^2|r,n
\rangle \geq \frac{1}{4}\left|\langle r,n|[a,a^\dagger]|r,n \rangle\right|^2,
\end{equation}
we also have
\begin{equation}
\langle r,n|(\triangle x)^2|r,n \rangle \langle r,n|(\triangle P)^2
|r,n \rangle \geq \left(\frac{1}{2} + \frac{p-1}{2} (-)^n \right)^2,
\end{equation}
which means that only for the parabose squeezed vaccum $|r,0 \rangle$,
the uncertainty relation reduces to an equality. Note that since
$[a,a^\dagger]$ is in general not a c-number, the right-hand side of (59)
itself depends on the given state. 
\par
In order to show the squeezing properties of the parabose squeezed number
states $|r,n \rangle$, let us consider the variances of the operators $x$ and
$P$ in the parabose number states
\begin{equation}
\langle n|(\triangle x)^2|n \rangle=\langle n|(\triangle P)^2|n \rangle
=n+ \frac{p}{2},
\end{equation}
which means that for $r\geq 0$, the states $|r,n \rangle$ are squeezed
in the $x$ direction 
\begin{eqnarray}
\langle r,n|(\triangle x)^2|r,n \rangle &\leq& \langle n|(\triangle x)^2
|n \rangle,   \nonumber\\
\langle r,n|(\triangle P)^2|r,n \rangle &\geq& \langle n|(\triangle P)^2
|n \rangle
\end{eqnarray}
and for $r<0$, the states $|r,n \rangle$ are squeezed in the $P$ direction 
\begin{eqnarray}
\langle r,n|(\triangle x)^2|r,n \rangle &\geq& \langle n|(\triangle x)^2
|n \rangle,   \nonumber\\
\langle r,n|(\triangle P)^2|r,n \rangle &\leq& \langle n|(\triangle P)^2
|n \rangle
\end{eqnarray}
respectively.

\section{Parabose Hermite polynomial states}
\par
In the previous section we show that the parabose squeezed vacuum
$|r,0 \rangle$ is a minimum uncertainty state for the normal squeezing. In
fact, $|r,0 \rangle$ is also a minimum uncertainty state for parabose
amplitude-squared squeezing. To see this, let us introduce two Hermitian
operators
\begin{equation}
Y_1 = \frac{1}{2} (a^2 + a^{\dagger 2}), ~~~
Y_2 = \frac{1}{2i} (a^2 - a^{\dagger 2}),
\end{equation}
which satisfy the commutator $[Y_1, Y_2]=2iN$ and the uncertainty relation
\begin{equation}
\langle \psi |(\triangle Y_1)^2|\psi \rangle \langle \psi|(\triangle Y_2)^2
|\psi \rangle \geq (\langle \psi|N|\psi \rangle)^2,
\end{equation}
where $N= \{a^{\dagger},a\}/2$ and $\langle \psi|(\triangle Y)^2 |\psi
\rangle$ stands for the variance of the operator $Y$ in the state
$|\psi \rangle$. It is easily to calculate the variances of $Y_1$ and $Y_2$ in
the parabose squeezed vacuum $|r,0 \rangle$
\begin{equation}
\langle r,0|(\triangle Y_1)^2|r,0 \rangle = \frac{p}{2}(cosh 2r)^2, ~~~
\langle r,0|(\triangle Y_2)^2|r,0 \rangle = \frac{p}{2}.
\end{equation}
Comparing with $\langle r,0|N|r,0 \rangle = \frac{p}{2} cosh 2r$, one finds
that
\begin{equation}
\langle r,0|(\triangle Y_1)^2|r,0 \rangle \geq \langle r,0|N|r,0 \rangle, ~~~
\langle r,0|(\triangle Y_2)^2|r,0 \rangle \leq \langle r,0|N|r,0 \rangle,
\end{equation}
which mean that the state $|r,0 \rangle$ is squeezed in the $Y_2$ direction.
\par
To search other minimum uncertainty states for parabose amplitude-squared
squeezing, following \cite{s9}, let us consider the eigenvalue eqaution
\begin{equation}
(Y_1 + i \lambda Y_2) |\psi \rangle = \beta |\psi \rangle,
\end{equation}
where $\lambda$ is real and $\beta$ is complex. Multiplying (68) by the
operator $Y_1 - i \lambda Y_2$ and then taking the inner product with $|\psi
\rangle$, one gets
\begin{equation}
\langle \psi|(\triangle Y_1)^2|\psi \rangle + \lambda^2 \langle \psi|
(\triangle Y_2)^2|\psi \rangle = 2\lambda \langle \psi|N|\psi \rangle.
\end{equation}
Again multiplying (68) by $Y_1 + i \lambda Y_2$ and taking the inner product
with $|\psi \rangle$ provides further information
\begin{equation}
\langle \psi|(\triangle Y_1)^2|\psi \rangle = \lambda^2 \langle \psi|
(\triangle Y_2)^2|\psi \rangle.
\end{equation}
>From these two equations we have
\begin{equation}
\langle \psi|(\triangle Y_1)^2|\psi \rangle = \lambda \langle \psi|N|\psi
\rangle, ~~~\langle \psi|(\triangle Y_2)^2|\psi \rangle = \frac{1}{\lambda}
\langle \psi|N|\psi \rangle.
\end{equation}
These equations show that a solution of (68) is a minimum uncertainty state
for parabose amplitude-squared squeezing, and the real $\lambda$ plays the
role of squeezing parameter.
\par
Now we begin to solve (68) for $|\psi \rangle$. First, in terms of creation
and annihilation operators, we can rewrite (68) as
\begin{equation}
(\frac{1-\lambda}{2}a^{\dagger 2} + \frac{1+\lambda}{2}a^2)|\psi \rangle =
\beta |\psi \rangle.
\end{equation}
Introducing a new state $|\psi^{\prime} \rangle = S(z)|\psi \rangle$, where
$S(z)=exp(\frac{z^{\ast}}{2}a^2 - \frac{z}{2} a^{\dagger 2})$ is the squeezed
operator and $z=re^{i\theta}$, one finds that $|\psi^{\prime} \rangle$
satisfies
\begin{eqnarray}
&& (\frac{1-\lambda}{2}cosh^2 r +\frac{1+\lambda}{2}e^{2i\theta} sinh^2 r)
a^{\dagger 2} |\psi^{\prime} \rangle \nonumber\\
&& + \frac{1}{2} (\frac{1-\lambda}{2}
e^{-i \theta} + \frac{1+\lambda}{2}e^{i \theta}) sinh 2r \{a^{\dagger},a \}
|\psi^{\prime} \rangle  \nonumber\\
&& +(\frac{1-\lambda}{2}e^{-2i\theta}sinh^2 r + \frac{1+\lambda}{2}cosh^2 r)
a^2 |\psi^{\prime} \rangle = \beta |\psi^{\prime} \rangle. 
\end{eqnarray}
Suitably choosing the parameter $z$, one can eliminate the coefficient of
$a^{\dagger 2}$ term. In doing so, for $0<\lambda<1$, we choose $\theta=\pi/2$
and $cosh r=\sqrt{\frac{1+\lambda}{2\lambda}}$, $sinh r=\sqrt{\frac{1-\lambda}
{2\lambda}}$; for $\lambda \geq 1$, we choose $\theta=0$, and $cosh r=\sqrt{
\frac{\lambda +1}{2}}$, $sinh r=\sqrt{\frac{\lambda -1}{2}}$.
Substituting these expressions into the above equation, for $0<\lambda<1$, we
have
\begin{equation}
(a^2 + \frac{i}{2}\sqrt{1-\lambda^2}\{a^{\dagger}, a\})|\psi^{\prime} \rangle
= \beta |\psi^{\prime} \rangle,
\end{equation}
and for $\lambda \geq 1$, we have
\begin{equation}
(a^2 + \frac{1}{2}\sqrt{\lambda^2 -1}\{a^{\dagger}, a\})|\psi^{\prime} \rangle
= \beta |\psi^{\prime} \rangle.
\end{equation}
We now expand $|\psi^{\prime} \rangle$ in terms of the parabose number states
(7), $|\psi^{\prime} \rangle = \sum_{n=0}^{\infty} c_n |n \rangle$, and
substitute this expression into (74) and (75). This leads to the recursion
relations $c_{n+2}=\frac{\beta-i\sqrt{1-\lambda^2}(n+p/2)}{\sqrt{[n+1][n+2]}}
c_n$ for $0<\lambda<1$, and $c_{n+2}=\frac{\beta -\sqrt{\lambda^2 -1}(n+p/2)}
{\lambda\sqrt{[n+1][n+2]}}c_n$ for $\lambda \geq 1$. Here we want to consider
a particularly simple subset of solutions. Noting that for $0<\lambda<1$, if
one takes $\beta= i\sqrt{1-\lambda^2}(m+\frac{p}{2})$, where $m$ is a
non-negative integer, then one has a truncated series for $|\psi^{\prime}
\rangle$ with $c_m|m \rangle$ being the last term. Similarly, for $\lambda
\geq 1$, and $\beta=\sqrt{\lambda^2 -1}(m+\frac{p}{2})$, one also gets a
polynomial expression for $|\psi^{\prime} \rangle$ with all $c_n=0$ when
$n>m$. Thus if $0<\lambda<1$, for even $m$, taking $c_1=0$, we have
\begin{equation}
|\psi^{\prime} (m,\lambda) \rangle = \sum_{k=0}^{\frac{m}{2}}
\frac{(i\sqrt{1-\lambda^2})^k 2^k (\frac{m}{2})!}
{\sqrt{[2k]!}(\frac{m}{2}-k)!} c_0 |2k \rangle,
\end{equation}
and for odd $m$, taking $c_0=0$, we have
\begin{equation}
|\psi^{\prime} (m,\lambda) \rangle = \sum_{k=0}^{\frac{m-1}{2}}
\frac{\sqrt{p}(i\sqrt{1-\lambda^2})^k 2^k (\frac{m-1}{2})!}
{\sqrt{[2k+1]!}(\frac{m-1}{2}-k)!} c_1 |2k+1 \rangle.
\end{equation}
If $\lambda \geq 1$, for even $m$, also taking $c_1=0$, we have
\begin{equation}
|\psi^{\prime} (m,\lambda) \rangle = \sum_{k=0}^{\frac{m}{2}}
\frac{(\sqrt{\lambda^2 -1})^k 2^k (\frac{m}{2})!}
{\lambda^k \sqrt{[2k]!}(\frac{m}{2}-k)!} c_0 |2k \rangle,
\end{equation}
and for odd $m$, taking $c_0=0$, we have
\begin{equation}
|\psi^{\prime} (m,\lambda) \rangle = \sum_{k=0}^{\frac{m-1}{2}}
\frac{\sqrt{p} (\sqrt{\lambda^2 -1})^k 2^k (\frac{m-1}{2})!}
{\lambda^k \sqrt{[2k+1]!} (\frac{m-1}{2}-k)!} c_1 |2k+1 \rangle.
\end{equation}
Using the deformed Hermite polynomial $H_m^{(p)}$, we can express these states
in a relatively compact form
\begin{equation}
|\psi^{\prime} (m,\lambda) \rangle = c_m (\lambda) H_m^{(p)} (i\gamma(\lambda)
a^{\dagger}) |0 \rangle,
\end{equation}
where $c_m (\lambda)$ is a normalization constant, and for $0<\lambda<1$,
$\gamma (\lambda)= e^{i\pi/4}\sqrt{\sqrt{1-\lambda^2}/2}$; for
$\lambda \geq 1$, $\gamma (\lambda)= \sqrt{\sqrt{\lambda^2 -1}/2\lambda}$.
Finally, combining this with the squeezing transformation we have the minimum
uncertainty states
\begin{equation}
|\psi (m,\lambda) \rangle = S^{-1} (z) |\psi^{\prime} (m,\lambda) \rangle
= c_m (\lambda) S^{-1} (z) H_m^{(p)} (i \gamma (\lambda) a^{\dagger})
|0 \rangle,
\end{equation}
where $z$ is chosen as mentioned earlier.
\par
We now want to examine some of the properties of the states $|\psi (m,\lambda)
\rangle$. Eq.(71) shows that if we know the average value of the operator $N$
in the state $|\psi (m,\lambda) \rangle$, then we know the variances of the
operators $Y_1$ and $Y_2$ in the same state. Using (54) and noticing that
$$
\langle 0|H_m^{(p)}(-i\gamma^{\ast} (\lambda) a) R H_m^{(p)} (i\gamma
(\lambda) a^{\dagger}) |0 \rangle= (-)^m/|c_m (\lambda)|^2,
$$
for $0<\lambda<1$ one can get
\begin{eqnarray}
\langle \psi (m,\lambda)|N|\psi (m,\lambda) \rangle &=&
\frac{1-\lambda^2}{\lambda} (m+\frac{p}{2}) + \frac{1+(p-1)(-)^m}{2} \lambda
\nonumber\\
& & + 2[m]^2 \lambda \sqrt{1-\lambda^2} \frac{|c_m (\lambda)|^2}
{|c_{m-1} (\lambda)|^2},
\end{eqnarray}
and for $\lambda \geq 1$ one can get
\begin{eqnarray}
\langle \psi (m,\lambda)|N|\psi (m,\lambda) \rangle &=&
\frac{\lambda^2 -1}{\lambda} (m+\frac{p}{2}) + \frac{1+(p-1)(-)^m}{2\lambda}
\nonumber\\
& & + 2[m]^2 \frac{\sqrt{\lambda^2 -1}}{\lambda^2} \frac{|c_m (\lambda)|^2}
{|c_{m-1} (\lambda)|^2}.
\end{eqnarray}
A straightforward calculation gives
$$
|c_0 (\lambda)|^{-2} =1,~~~ |c_1 (\lambda)|^{-2}= 4[1] |\gamma|^2,
$$
$$
|c_2 (\lambda)|^{-2}=16[2]!|\gamma|^4 + ([2]!)^2.
$$
Let us consider the state $|\psi (1,\lambda)
\rangle$. We find for $0<\lambda<1$ that
\begin{eqnarray}
&& \langle \psi (1,\lambda)|N|\psi (1,\lambda) \rangle = \frac{2+p}{2\lambda},
\nonumber\\
&& \langle \psi (1,\lambda)|(\triangle Y_1)^2|\psi (1,\lambda) \rangle
=1+\frac{p}{2}, \nonumber\\
&& \langle \psi (1,\lambda)|(\triangle Y_2)^2|\psi(1,\lambda)
\rangle = \frac{2+p}{2\lambda^2},
\end{eqnarray}
and for $\lambda \geq 1$ that
\begin{eqnarray}
&& \langle \psi (1,\lambda)|N|\psi (1,\lambda) \rangle= \frac{2+p}{2}\lambda,
\nonumber\\
&& \langle \psi (1,\lambda)|(\triangle Y_1)^2|\psi (1,\lambda) \rangle
=\frac{2+p}{2}\lambda^2 \nonumber\\
&& \langle \psi (1,\lambda)|(\triangle Y_2)^2|
\psi(1,\lambda) \rangle = 1+\frac{p}{2},
\end{eqnarray}
It is obviously that as $\lambda \rightarrow 0$, $Y_1$ becomes increasingly
squeezed, and $\langle \psi (1,\lambda)|(\triangle Y_1)^2|\psi (1,\lambda)
\rangle$ is equal to $1+\frac{p}{2}$. Similarly, as $\lambda \rightarrow
\infty$, $Y_2$ becomes more and more squeezed and $\langle \psi (1,\lambda)
|(\triangle Y_2)^2|\psi (1,\lambda) \rangle$ is equal to $1+\frac{p}{2}$. Then
we examine the state $|\psi (2,\lambda) \rangle$. For $0<\lambda<1$, we have
\begin{equation}
\langle \psi (2,\lambda)|N|\psi (2,\lambda) \rangle = \frac{(2+p)(4+p)-(8+6p)
\lambda^2}{2\lambda (2+p-2\lambda^2)},
\end{equation}
and for $\lambda \geq 1$, we have
\begin{equation}
\langle \psi (2,\lambda)|N|\psi (2,\lambda) \rangle =
\frac{(2+p)(4+p)\lambda^3 -(8+6p)\lambda}{2(2+p)\lambda -4}.
\end{equation}
In this case, however, as $\lambda$ goes to zero, $\langle \psi(2,\lambda)|
(\triangle Y-1)^2|\psi(2,\lambda) \rangle$ goes to $2+\frac{p}{2}$, and as
$\lambda$ goes to infinity, $\langle \psi(2,\lambda)|(\triangle Y_2)^2|\psi
(2,\lambda) \rangle$ goes to $2+\frac{p}{2}$. Generally, for any fixed
non-negative integer $m$, since when $\lambda \rightarrow 0$ or $\lambda
\rightarrow \infty$, $|c_m(\lambda)|^2$ goes to a number which only depends on
the given integer $m$ and the parastatistica order $p$, we have
$$
\lim_{\lambda \rightarrow 0} \langle \psi(m,\lambda)|(\triangle Y_1)^2|\psi
(m,\lambda) \rangle = m+\frac{p}{2},
$$
and
$$
\lim_{\lambda \rightarrow \infty} \langle \psi(m,\lambda)|(\triangle Y_2)^2
|\psi (m,\lambda) \rangle = m+\frac{p}{2}.
$$
A short calculation shows that the state $|\psi (1,\lambda) \rangle$ is not
squeezed in the normal sense.
\par
In summary, we have introduced a new kind of deformation of the usual Hermite
polynomials and discussed their main preperties in this paper. These deformed
Hermite polynomials may have some applications in studying parabose systems.
For instance, they can be used to give the explicit form for the parabose
squeezed number states, as well as to present a particularly simple subset of
the minimum uncertainty states for parabose amplitude-squared squeezing. It is
reasonable to believe that the para-deformed Hermite polynomials have other
applications in parastatistics, and this is the task of oue next step.

\end{document}